\documentclass[prd,preprint,superscriptaddress]{revtex4}
\usepackage{revsymb}
\usepackage{amsmath}
\usepackage{graphicx}
\usepackage{bm}
\newcommand{\be}{\begin{equation}}
\newcommand{\ee}{\end{equation}}
\newcommand{\ben}{\begin{eqnarray}}
\newcommand{\een}{\end{eqnarray}}

\newcommand{\vphi}{\varphi}
\begin{document}
\title{Soft coincidence in late acceleration}
\author{Sergio del Campo\footnote{Electronic Mail-address:
sdelcamp@ucv.cl}} \affiliation{Instituto de F\'{\i}sica,
Pontificia Universidad Cat\'{o}lica de Valpara\'{\i}so, Avenida
Brasil 2950, Casilla 4059, Valpara\'{\i}so, Chile}
\author{Ram\'{o}n Herrera\footnote{E-mail address: rherrera@unab.cl}}
\affiliation{Departamento de Ciencias F\'{\i}sicas, Universidad
Andr\'{e}s Bello, Av. Rep\'{u}blica 273, Santiago, Chile}
\author{Diego Pav\'{o}n\footnote{E-mail address: diego.pavon@uab.es}}
\affiliation{Departamento de F\'{\i}sica, Facultad de Ciencias,
Universidad Aut\'{o}noma de Barcelona, 08193 Bellaterra (Barcelona), Spain}

\begin{abstract}
We study the coincidence problem of late cosmic acceleration by
assuming that the present ratio between dark matter and dark
energy is a slowly varying function of the scale factor. As dark
energy component we consider two different candidates, first a
quintessence scalar field, and then a tachyon field. In either
cases analytical solutions for the scale factor, the field and the
potential are derived. Both models show a good fit to the recent
magnitude-redshift supernovae data. However, the likelihood
contours disfavor the tachyon field model as it seems to prefer a
excessively high value for the matter component.
\end{abstract}

\maketitle

\section{Introduction}
Nowadays it is widely accepted that the present stage of cosmic
expansion is accelerated \cite{reviews,adam} albeit there are
rather divergent proposals about the mechanism behind this
acceleration. A cosmological model of present acceleration should
not only fit the high redshift supernovae data, the cosmic
microwave background anisotropy spectrum  and safely pass other
tests, it must solve the coincidence problem as well, namely ``why
the Universe is accelerating just now?", or in the realm of
Einstein gravity ``why are the densities of matter and dark energy
of precisely the same order today?" \cite{coincidence} -note that
these two energies scale differently with redshift. While it might
happen that this coincidence is just a ``coincidence" -and as such
no explanation is to be found- we believe models that fail to
account for this cannot be regarded as satisfactory.

In a class of models designed to solve this problem  the dark
energy density ``tracks" the matter energy density for most of the
history of the Universe, and overcomes it only recently (see,
e.g., Ref. \cite{trackers}). However, these models suffer the
drawback of fine-tuning the initial conditions whereupon they are
not, after all, much better than the conventional ``concordance"
model which rests on a mixture of matter and a fine-tuned
cosmological constant \cite{sabino}.

There is an especially successful subset of models based on an
interaction between dark energy and cold matter (i.e., dust) such
that the ratio $r$ of the corresponding energy densities tends to
a constant of order unity at late times
\cite{plb,luca,scale,interacting,dw,somasri,olivares} thus solving
the problem. However, the current observational information does
not necessarily imply that $r$ ought to be  strictly constant
today. For the coincidence problem to be addressed a softer
condition may suffice, namely that at present $r$ should be a
slowly varying function of the scale factor with $r(a= a_{0})
\simeq 3/7$, the currently observed ratio. By {\em slowly varying}
we mean that the current rate of variation of  $r(a)$ should be no
much larger than $H_{0}$, where $H \equiv \dot{a}/a$ denotes the
Hubble factor of the Friedamnn-Lemaitre-Robertson-Walker (FLRW)
metric, and a zero subscript means present time. It should be
noted that because the nature of dark matter and dark energy is
largely unknown an interaction between both cannot be excluded a
priori. In fact, the possibility has been suggested from a variety
of angles \cite{angles}.

To avoid a possible conflict with observational constraints on
long-range forces \cite{only} we consider that the baryon
component of the matter does not participate in the the
interaction and, further, to simplify the analysis -i.e., in order
not to have an uncoupled component- we exclude the baryons
altogether. While this might be seen as a radical step it should
be taken into account that our study restricts itself to times
near the present time and these are characterized, among other
things, by a low value of the baryon energy density ($5$\% or less
of the total energy budget, approximately six times lower than the
dark matter contribution and fourteen below the dark energy
component \cite{spergel}) whereby it should not significantly
affect our results. This is in keeping with the findings of
Majerotto {\em et al.} \cite{majerotto}. For interacting models
encompassing most the Universe history in which the baryons enter
the dynamical equations as a non-interacting component, see Refs.
\cite{majerotto} and \cite{luca}.

The target of this paper is to present two models of late
acceleration that fulfill ``soft coincidence", namely: $(i)$ when
the dark energy is a quintessence scalar field, and $(ii)$ when
the dark energy is a tachyon field. The latter was introduced by
Sen \cite{sen} and soon afterwards it became  a candidate for
driving inflation as well as late acceleration -see e.g., Ref.
\cite{gary}. The outline of this paper is as follows: Section II
considers the quintessence model with a constant equation of state
parameter. There it is assumed that the quintessence field slowly
decays into dark matter with the equation of state of dust.
Section III considers the tachyon field and again assumes a slowly
decay into dust. This time, however, the equation of state
parameter is allowed to vary. Finally, section IV summarizes our
findings.

\section{The quintessence interacting model}
We consider  a two-component system, namely, cold dark matter,
described by an energy density $\rho_{m}$, and a quintessence
scalar field $\phi$ with energy density and pressure defined by
\\
\begin{equation}
\rho_\phi = \textstyle{1\over{2}}\dot{\phi}^2 + V(\phi)\, ,
\qquad \mbox{and} \qquad
p_{\phi}  = \textstyle{1\over{2}}\ \dot{\phi}^2 - V(\phi),
\label{rhop1}
\end{equation}
\\
respectively, in a spatially flat FLRW universe. The over dot
indicates derivative with respect to the cosmic time and $V(\phi)$
is the quintessence scalar potential. We assume that these two
components do not evolve separately but interact through a source
(loss) term (say, $Q$) that enters the energy balances
\\
\begin{equation}
\dot{\rho}_m + 3 H \rho_{m}  = Q
\label{cons1}
\end{equation}
\\
and
\\
\begin{equation}
\dot{\rho}_{\phi} + 3 H (\rho_{\phi} + p_{\phi}) = -Q ,
\label{cons2}
\end{equation}
\\
where in view of (\ref{rhop1}) last equation is equivalent to
\\
\begin{equation}
\dot{\phi} \left[\ddot{\phi} + 3 H \dot{\phi} + V' \right] = - Q .
\label{cons3}
\end{equation}
\\

In the following we constrain the interaction $Q$ by demanding
that the solution to Eqs. (\ref{cons1}) and (\ref{cons2}) be
compatible with a variable ratio between the energy densities
$r(x)\equiv \rho_{m}/\rho_{\phi}$, where $x = a/a_{0}$ is the
normalized scale factor, and that around the present time $r(x)$
is a smooth, nearly constant function with $r(x=1) = r_{0}$ being
of order one. We also assume that the quintessence component obeys
a barotropic equation of state, it is to say $p_{\phi} = w_{\phi}
\rho_{\phi}$ with $w_{\phi}$ a negative constant (a distinguishing
feature of dark energy fields -quintessence fields or whatever- is
a high negative pressure). In virtue of these relations the set of
dynamical equations reduces to a single equation
\\
\begin{equation}
\dot{\rho}_{\phi} + 3 H \left( 1 + \frac{w_{\phi}+ \frac{ \dot{r}}{3 H}}{1 +r} \right)\rho_{\phi}  = 0 \, ,
\label{cons6}
\end{equation}
\\
whose solution is
\\
\begin{equation}
\rho_{\phi}(x) =  \rho_{\phi}^{(0)} \, e^{-3I(x)},
\label{rhos1}
\end{equation}
\\
with
\\
\begin{equation}
I(x) =  \int_1^x{F(x') \frac{d x'}{x'}}\, , \quad \mbox{where}\, \quad
F(x) = 1+ \frac{w_{\phi} + {\textstyle{1\over3}}x r'(x)}{1 +r(x)}\, ,
\label{int1}
\end{equation}
\\
and the prime means derivation with respect to $x$.
On the other hand, by combining Friedmann's equation
\\
\begin{equation}
3 H^2 = \kappa (\rho_{m} + \rho_{\phi})  \qquad(\kappa \equiv 8\pi G)
\label{Fried1}
\end{equation}
\\
with Eq. (\ref{rhos1}) we get
\\
\begin{equation}
H(x) = H_{0} \, \sqrt{\frac{1+ r(x)}{1 + r_{0}}} \, e^{-\frac{3}{2}I(x)},
\label{hx1}
\end{equation}
\\
where $H_{0} = \sqrt{\rho_{\phi}^{(0)}  \kappa /3 }$ denotes the
current value of the Hubble factor. From this, it follows that
\\
\begin{equation}
\frac{H_{0}}{\sqrt{1 + r_{0}}}\left[ t(x)-t_{0}\right] = \int_1^x
{\frac{e^{\frac{3}{2}I(x')}}{\sqrt{1 + r(x')}}\frac{d x'}{x'}}.
\label{ta1}
\end{equation}
\\
If this integral could be solved analytically, we would obtain the
scale factor in terms of the cosmological time.

Equations (\ref{cons2}) and (\ref{cons6}) alongside
~(\ref{rhos1}), ~(\ref{int1}), and ~(\ref{hx1}) imply
\\
\begin{equation}
Q(x) = \frac{3 H_0 \rho_{\phi}^{(0)}}{\sqrt{1 +
r_{0}}}\,\frac{1}{\sqrt{1 + r(x)}}\,\left[\frac{x\,r'(x)}{3}
-w_{\phi}\,r(x)\right]\, e^{-\textstyle{9\over{2}} I(x)}.
\label{del3}
\end{equation}
\\
Note that $Q(x)$ is a positive-semidefinite function, as it
should. A negative $Q(x)$ would imply a transfer of energy from
the matter to the scalar field which migth violate the second law
of thermodynamics. While in view of the unknown nature of dark
matter and dark energy we cannot say for certain that these
components fulfill the aforesaid law, in the absence of any
evidence against it, the most natural thing is to assume that they
obey it.

From the definitions~(\ref{rhop1}) and the equation of state
$p_{\phi} = w_{\phi}\rho_{\phi}$ the quintessence field  and its
potential are given by
\\
\begin{equation}
\phi(x) = \phi_0 + \sqrt{\frac{3(1+w_{\phi})}{\kappa}}\int_1^x {
\sqrt{\frac{1} {1 + r(x')}} \, \frac{dx'}{x'}}, \label{sQ1}
\end{equation}
\\
and
\\
\begin{equation}
V(x) = V_{0}\, e^{-3I(x)},
\label{VQ1}
\end{equation}
\\
respectively. Here $\phi_{0}$ is an integration constant, and
$V_{0} = \textstyle{1\over{2}}\left(1 - w_{\phi} \right)\rho_{\phi}^{(0)} $.\\

As said before, we apply the above formalism to the case in which
the variable $x$ is not far from unity whence the ratio $r(x)$ can
be approximated by
\\
\begin{equation}
r(x) \simeq r_{0} + \varepsilon_{0}(1 - x), \label{eta1}
\end{equation}
\\
where $r_{0}$ is the present value of the ratio between the energy
densities $\rho_{m}$ and $\rho_{\phi}$, and $\varepsilon_{0}$ is a
small positive-definite constant. We do not consider negative
values for $\varepsilon_{0}$ since it would imply that $r(x)$ was
increasing in the recent past and therefore that it oscillates.
While we are unaware of any definitive argument against this
possibility it looks certain that only contrived models may lead
to this behavior. Further, oscillations in $r(x)$ may seriously
jeopardize the well tested picture of structure formation
\cite{springel}. (Note in passing  that the choice (\ref{eta1})
implies that $\mid \dot{r}(x)\mid \lesssim H_{0}$ for $x \sim
{\cal O}(1)$).

It follows that
\\
\begin{equation}
F(x) =1 + \frac{\alpha_{1} - \alpha_{2} x}{\alpha_{3} - \alpha_{4}
x} \, , \label{eta2}
\end{equation}
\\
where the $\alpha_{i}$ are constants given by
\\
\begin{equation}
\alpha_{1} = w_{\phi}, \quad \alpha_{2} =
\textstyle{1\over{3}}\varepsilon_{0}, \quad \alpha_{3} = 1 + r_{0}
+ \varepsilon_{0}, \quad \alpha_4 = \varepsilon_{0},
\label{alphai}
\end{equation}
\\
respectively.
With this, we obtain
 \\
\begin{equation}
\rho_{m}(x) = \rho_{\phi}^{(0)} \left(\frac{\alpha_{3} - \alpha_{4}}{\alpha_{3} - \alpha_{4} x }\right)
(\alpha_{3}-1-\alpha_{4}\,x) \,x^{-3\left(\frac{\alpha_{1} + \alpha_{3}}{\alpha_{3} }\right)}.
\label{rhom3}
\end{equation}
\\
and
\\
\begin{equation}
\rho_{\phi}(x) = \rho_{\phi}^{(0)} \left(\frac{\alpha_{3} - \alpha_{4}}{\alpha_{3} - \alpha_{4} x }\right)
\,x^{-3\left(\frac{\alpha_{1}+\alpha_{3}}{\alpha_{3} }\right)} ,
\label{rhos3}
\end{equation}

The Hubble parameter can be written as
\\
\begin{equation}
H(x) = H_{0} \,x^{- \frac{3}{2}\left(\frac{\alpha_1  + \alpha_3}{\alpha_3 }\right)} ,
\label{hx3}
\end{equation}
\\
where $H_{0}$ was given above. From this expression the scale
factor is shown to follow a power-law dependence on time
\\
\begin{equation}
a(t) = a_{0} \left(\frac{t}{t_{0}}\right)^{\frac{2\,\alpha_{3}}{3 (\alpha_{1} + \alpha_{3})}}.
\label{at1}
\end{equation}
\\

For the Universe to accelerate, the constraint $\frac{2\,\alpha_3}{
3 \left(\alpha_1 + \alpha_3 \right)} >1$ must be fulfilled, i.e.,
\\
\begin{equation}
1+ r_{0}+ \varepsilon_{0} < -3 w_{\phi}. \label{rest1}
\end{equation}
\\
This, alongside the condition that the energy densities decrease with expansion implies
\\
\begin{equation}
-w_{\phi} < 1+r_{0}+\varepsilon_{0} < -3w_{\phi}. \label{rest2}
\end{equation}

\begin{figure}[th]
\includegraphics[height=5.5in,width=6.5in,angle=0,clip=true]{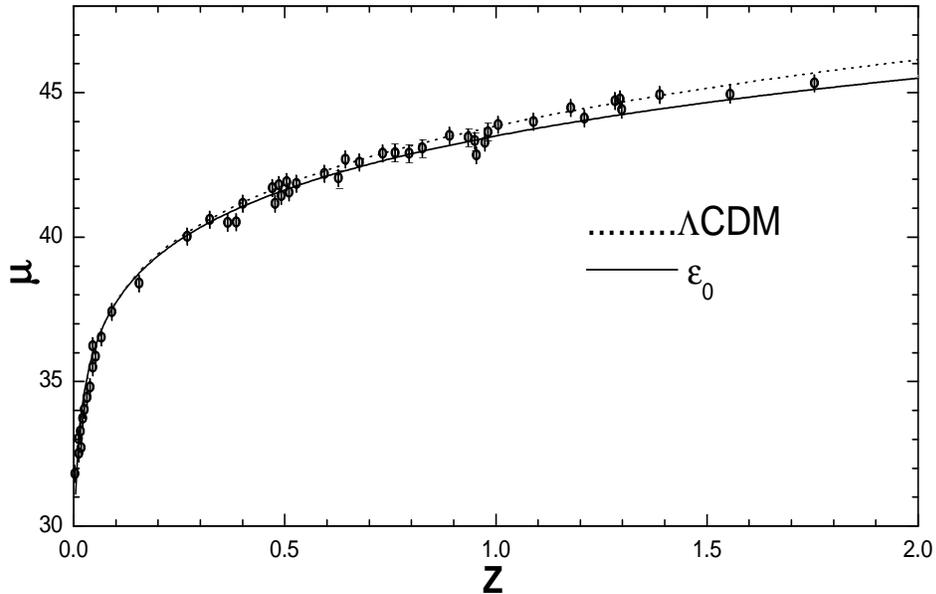}
\vspace{-5.0cm}\caption{Distance moduli vs redshift for the
quintessence--dark matter interacting model. In plotting the
graphs the expression $ \mu = 5 \log d_{L} +25$, with $d_{L} =
(1+z) \int^{z}_{0}H^{-1}(z') dz'$ in units of megaparsecs was
assumed. Here $w_{\phi}=-0.75$, $\varepsilon_{0} = 0.077$ and
$r_{0}= 0.27$. For comparison we have also plotted the prediction
of the concordance $\Lambda$CDM model with $\Omega_{m0} = 0.3$.
The data points correspond to the ``gold" sample of type Ia
supernovae of Ref. \cite{adam}.} \label{fisupernova}
\end{figure}

Fig. \ref{fisupernova} shows a good fit to the ``gold"  set of
supernovae data points of Riess {\it et al.} \cite{adam}.
\\

\begin{figure}[th]
\includegraphics[height= 6in,width=7in,angle=0,clip=true]{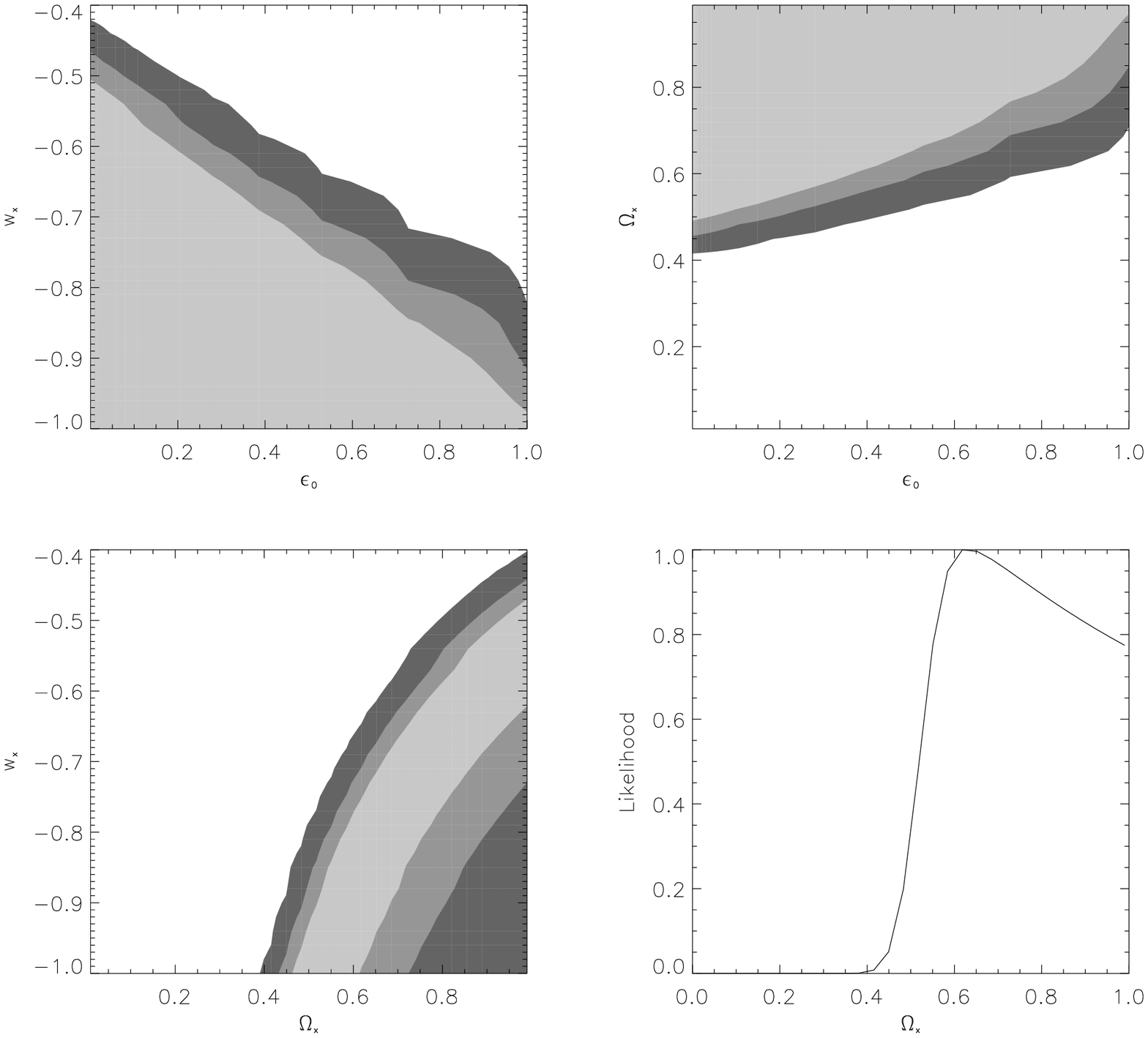}
\vspace{0cm}\caption{Likelihood contours for the
quintessence--matter interacting model showing the 68\%, 90\% and
99\% confidence intervals. The likelihoods are marginalized over
the rest of parameters. The prior $\Omega_{m} + \Omega_{\phi} = 1$
was used. The right bottom panel shows the probability function of
the quintessence parameter density.} \label{likesoft1}
\end{figure}

The likelihood contours are depicted in Fig. \ref{likesoft1}. The
mean values of the free parameters are: $\Omega_{\phi} = 0.78538$,
$w_{\phi} = -0.757284$, $\varepsilon_{0} = 0.0777764$. Notice that
$\Omega_{\phi}$ is above the concordance $\Lambda$CDM value of
$0.73$ and that $w_{\phi}$ is significantly larger than the value
found in non-interacting models. However, we have not considered
phantom fields (scalar fields as given by Eq. (\ref{rhop1}) do not
encompass phantom behavior), otherwise a shift of $w_{\phi}$
toward more negative values should be expected. Notice (top
panels) that the parameter $\varepsilon_{0}$ is rather degenerate.

\begin{figure}[th]
\includegraphics[width=5.0in,angle=0,clip=true]{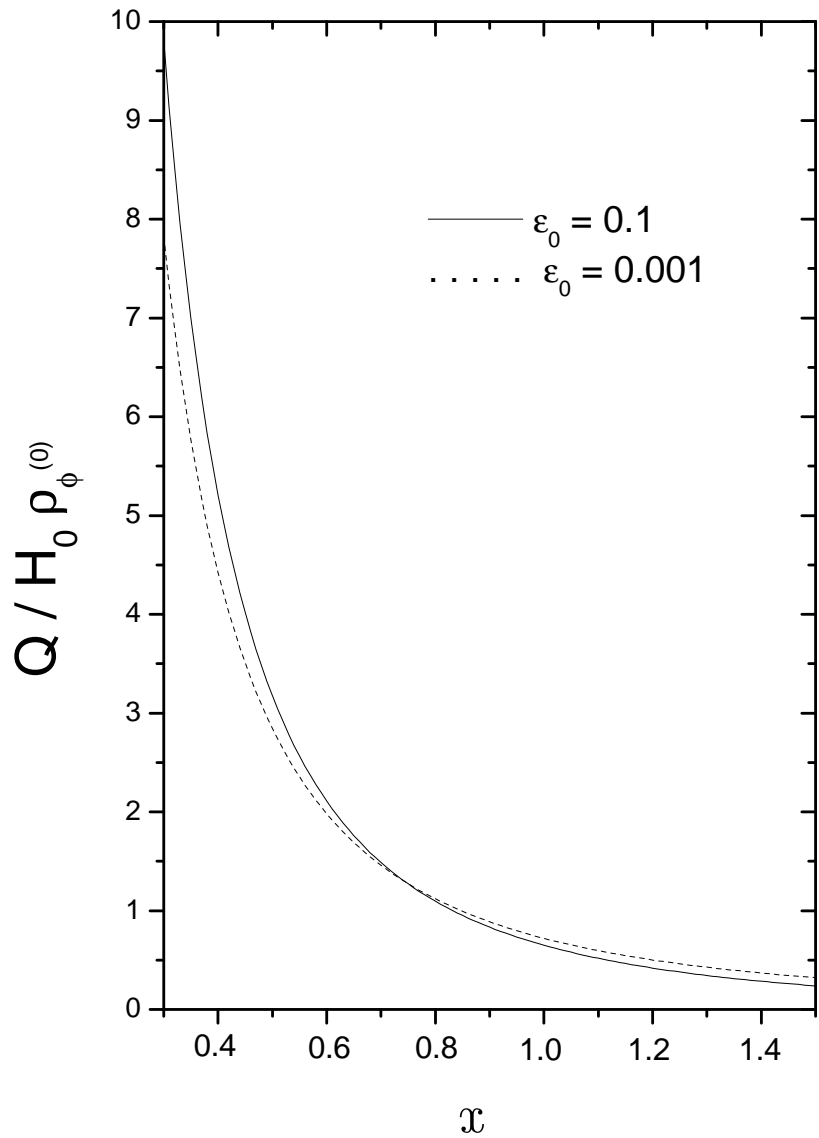}
\caption{Evolution of the interaction term with expansion. The
lower $\varepsilon_{0}$, the slower the decrease of the
quintessence field. Here $w_{\phi}= -0.8$ and $r_0 =3/7$}.
\label{qfi}
\end{figure}

Under restriction ~(\ref{rest1}) the interaction term reads
\\
\begin{equation}
Q(x) = -3 H_{0}\rho_{\phi}^{(0)}\left[w_{\phi}+\frac{\alpha_{4}
x}{3[(\alpha_{3} - 1-\alpha_{4} x)]}\right]
(\alpha_{3}-\alpha_{4})\frac{\left( \alpha_3 -1 -\alpha_4 x
\right)}{\left(\alpha_3 - \alpha_4 x \right)^2}
\,x^{-\textstyle{9\over{2}}\left(\frac{\alpha_{1}  +
\alpha_{3}}{\alpha_{3} }\right)}, \label{delta2}
\end{equation}
\\
see Fig. \ref{qfi}.
\\

\begin{figure}[th]
\includegraphics[width=5.0in,angle=0,clip=true]{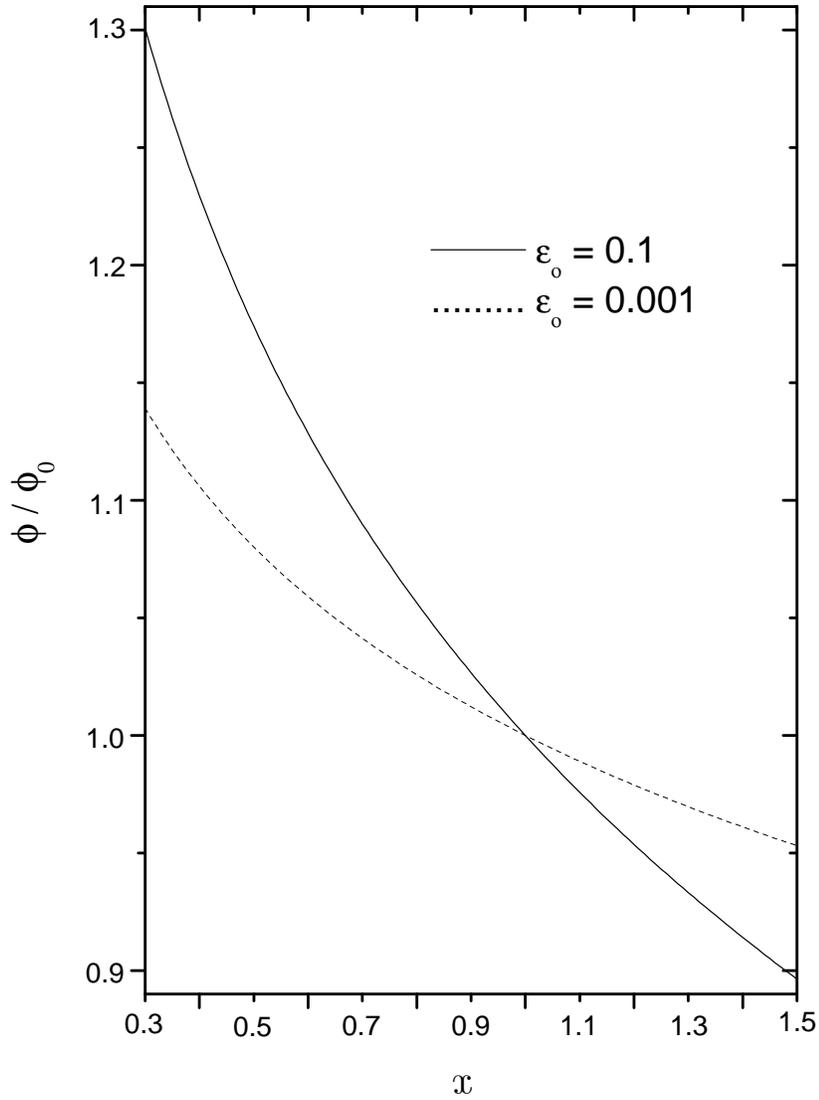}
\caption{Evolution of the scalar field with expansion. Again,
$w_{\phi}= -0.8$ and $r_0= 3/7$.} \label{fi}
\end{figure}

In its turn, the scalar field $\phi$ and the scalar potential $V(x)$ are given by
 \\
\begin{equation}
\phi(x) = \phi_{0} \left[ \frac{\tanh^{-1}\left(\sqrt{1 - \frac{\alpha_{4}}{\alpha_{3}} x} \right)}
{\tanh^{-1}\left(\sqrt{1 - \frac{\alpha_4}{\alpha_{3}}}\right)}\right] \, ,
\label{qx2}
\end{equation}
\\
and
 \\
\begin{equation}
V(x) = V_{0} \left( \frac{\alpha_{3} - \alpha_{4}}{\alpha_{3} - \alpha_{4} x} \right)
\,x^{-3\left(\frac{\alpha_{1}  + \alpha_{3}}{\alpha_{3} }\right)},
\label{Vx1}
\end{equation}
\\
respectively. Here $\phi_0 = \sqrt{\frac{12}{\kappa}\left(\frac{1+
w_{\phi}}{\alpha_3} \right)} \tanh^{-1}\left[\sqrt{1 -
\frac{\alpha_4}{\alpha_3}} \right]$, and $V_{0} =
\textstyle{1\over{2}}(1 - w_{\phi}) \rho_{\phi}^{(0)}$, see Fig.
\ref{fi}.
\\
\begin{figure}[th]
\includegraphics[width=5.0in,angle=0,clip=true]{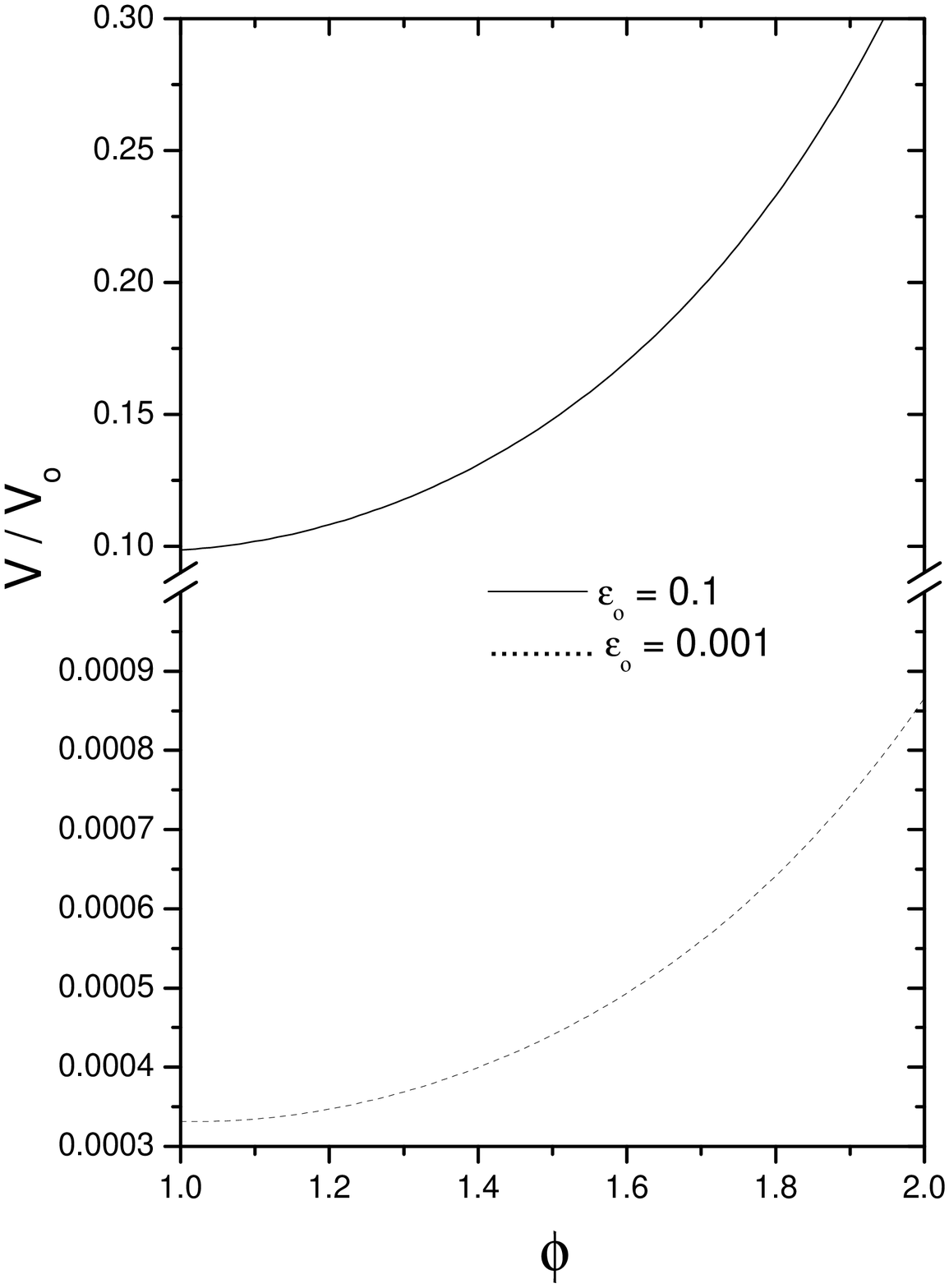}
\caption{ The effective potential versus the scalar field. Here
$w_{\phi}= -0.8$ and $r_0= 3/7$.} \label{potfi}
\end{figure}

Equations ~(\ref{qx2}) and~(\ref{Vx1}) lead to
 \\
\begin{equation}
V(\phi) =  \frac{ V_{0}\left(1 -
\frac{\alpha_{4}}{\alpha_{3}}\right)}{\tanh^{2} \left(
\sqrt{\frac{\kappa \alpha_{3}}{12 \left(1 + w_{\phi}\right)}}\,
\phi \right)} \,\left[\sqrt{\frac{\alpha_{3}}{\alpha_{4}}} \, \,
\mbox{sech} \left( \sqrt{\frac{\kappa \alpha_3} {12 \left(1 +
w_{\phi}\right)}}\,\phi \right)
 \right]^{-6\left(\frac{\alpha_{1} + \alpha_{3}}{\alpha_{3}}\right)}\, .
\label{Vx3}
\end{equation}
\\
Figures \ref{fi} and \ref{potfi} taken together  show that the
potential decreases with the Universe expansion. While we do not
know whether there is any field theory backing this potential it
is intriguing to see that around $\phi = 0$ it behaves  as
\\
\begin{equation}
V(\phi)  \sim C_{1} \phi^{-2}+ C_{2} + C_{3} \phi^{2} + C_{4} \phi^{4} + ... \,
\label{vexp}
\end{equation}
\\
where the $C_{i}$'s are constants. The first term is used in
quitessence models -see e.g., Ref. \cite{potentials}, whereas the
third and fourth terms of the expansion are  well--known
potentials in inflation theory (chaotic potentials) \cite{Linde};
the second term plays the role of a cosmological constant. This
leads us to surmise that, in reality, $V(\phi)$ might be
considered an effective potential resulting from the combination
of a number of fields.

\section{The tachyon interacting model}
The tachyon field naturally emerges as a straightforward
generalization of the Lagrangian of the relativistic particle much
in the same way as the scalar field $\phi$ arises from
generalizing the Lagrangian of the non--relativistic particle
\cite{bagla}. Recently, the realization of its potentiality as
dark matter \cite{shiu} and dark energy \cite{awakened} has
awakened the interest in it. We begin by succinctly recalling the
basic equations of the tachyon field to be used below where its
interaction with cold dark matter (dust) will be considered.

The stress--energy tensor of the tachyon field
\\
\begin{equation}
T^{(\vphi)}_{ab} = \frac{V(\vphi)}
{\sqrt{1+{\vphi}^{,c} {\vphi}_{,c}}}
\left[-g_{ab}\left(1 +{\vphi}^{,c} {\vphi}_{,c} \right) +
\vphi_{,a}\vphi_{,b}\right]
\ ,
\label{stress1}
\end{equation}
\\
admits to be written in the perfect fluid form
\\
\begin{equation}
T^{(\vphi)}_{ab} = \rho_{\vphi} u_{a}u_{b} + p_{\vphi} \left(g_{ab} +
u_{a}u_{b}\right),
\label{stress2}
\end{equation}
\\
where the energy density and pressure are given by
\\
\begin{equation}
\rho_{\vphi}  = \frac{V(\vphi)}{\sqrt{1-\dot{\vphi}^2}}\, ,
\qquad \mbox{and} \qquad
p_{\vphi}  = - V(\phi){\sqrt{1-\dot{\vphi}^2}},
\
\label{rhop}
\end{equation}
\\
respectively, with
\\
\begin{equation}
\dot{\vphi } \equiv  \vphi_{,a}u^{a}
= \sqrt{-g ^{ab}\vphi_{,a}\vphi_{,b}}\, , \quad \mbox{and}
\quad u_{a} = - \frac{\vphi_{,a}} {\dot{\vphi}}\ ,
\quad \mbox{where} \quad
u ^{a}u _{a} = -1.
\label{respectively}
\end{equation}

In the absence of interactions other than gravity the evolution of
the energy density is governed by $\dot{\rho}_{\vphi}= - 3
H\dot{\vphi}^2\rho_{\vphi}$, therefore when $\dot{\vphi}^2 < 1$ it
decays at a lower rate than that for dust. It approaches the
behavior of dust for $\dot{\vphi}^2 \rightarrow 1$, thereby in
this limit the tachyon field behaves dynamically as pressureless
matter does.  Consequently, we shall assume $\dot{\vphi}^2 < 1$
since for $\dot{\vphi}^2 = 1$ both components obey the same
equation of state for dust.

For an interacting mixture of a tachyon field and cold dark matter, with energy density
$\rho_{m}$ and negligible pressure, the interaction term, $Q$, between
these two components is described by the following balance equations
\\
\begin{equation}
\dot{\rho}_{m} + 3H (\rho_{m}+ \Pi_{m}) = Q \ , \label{balance1}
\end{equation}
\begin{equation}
\dot{\rho}_{\vphi} + 3H \dot{\vphi}^2 \rho_{\vphi} = -Q \, .
\label{balance2}
\end{equation}
\\
 The $\Pi_{m}$ term, on the left hand side
of Eq. (\ref{balance1}), accounts for the fact that the matter
component may be endowed with a viscous pressure or perhaps it is
slowly decaying into dark matter and/or radiation \cite{decay}. In
either case one can model this term as $\Pi_{m} = \alpha \rho_{m}
H$ with $\alpha$ a small negative constant since $\Pi_{m}$ is a
small correction to the matter pressure -see \cite{winfried} and
references therein.

As before, we consider the ratio between the densities of matter
and tachyonic energy a function $r(x)$ of the normalized scale
factor (to be specified later), and  again, we must have $Q(x)
>0$ -its expression is to be found below. Then,
equations ~(\ref{balance1}) and~(\ref{balance2}) combine to
\\
\begin{equation}
\dot{\rho}_{\vphi} + 3H \left[1+\frac{w_\vphi + \alpha r(x)
+\frac{r'(x) \, x}{3}} {1+ r(x)} \right]\rho_{\vphi} = 0   \qquad
(x \equiv a/a_{0})\, .
 \label{16}
\end{equation}
\\
The latter can be solved to
\\
\begin{equation}
\rho_\vphi(x) = \rho_\vphi^{(0)} e^{-3\tilde{I}(x)}\, , \qquad
\mbox{where} \qquad  \tilde{I}(x) =  \int_1^x{\tilde{F}(x')\,
\frac{d x'}{x'}}, \label{15}
\end{equation}
\\
with
\\
\begin{equation}
\tilde{F}(x) = 1+ \frac{w_\vphi(x) + \alpha r(x) + \textstyle{1\over{3}}r'(x)x}{1 + r(x)}.
\label{fx2}
\end{equation}

The interaction term takes the form
\\
\begin{equation}
Q(x) =  3\rho_\vphi^{(0)} H(x) \left(\frac{r(x)}{1+
r(x)}\right)\left[\alpha -w_\vphi(x) + \frac{x r'(x)}{
3 r(x)}\right ] e^{-3 \tilde{I}(x)}.
\label{delta3}
\end{equation}
\\
and the tachyonic scalar field and its potential obey
\\
\begin{equation}
\vphi(x) = \vphi_0 + \frac{\sqrt{1 + r_{0}}}{H_0} \int_1^x
{\sqrt{\frac{1 + w_\vphi (x')}{1 + r(x')}}\, e^{\frac{3}{2}\tilde{I}(x')}} \, \frac{d x'}{x'},
\label{vphix1}
\end{equation}
\\
and
\\
\begin{equation}
V(x) = V_{0} \sqrt{-w_\vphi (x)}\,\,e^{-3 \tilde{I}(x)},
\label{Vvphix1}
\end{equation}
\\
respectively.

Up to now we have left the ratio function $r(x)$ free. As before,
we specify it for $x$ values around unity as
\\
\begin{equation}
r(x) \simeq r_{0} + \varepsilon_{0}(1 - x),
\label{eta3}
\end{equation}
\\
where $r_{0} = (\rho_{m}/\rho_{\vphi})_{0}$, and $\varepsilon_{0}$
is once again  a small positive-definite constant. Likewise, we
assume that the equation of state parameter $w_{\vphi}$ is given
by
\\
\begin{equation}
w_{\vphi}(x) = w_{0} + w_{1}(1 - x),
\label{eta4}
\end{equation}
\\
where $w_{0}$ and $w_{1}$ are constants, the first one denotes the
current value of the $w_{\vphi}(x)$ function,  and the second one
is minus its first derivative, which is expected to be small. Thus,
\\
\begin{equation}
\tilde{F}(x) =1 + \frac{a_1 - b_1 x}{a_2 - b_2 x},
\label{tilFx}
\end{equation}
\\
where the constants $a_{i}$ and $b_{i}$ stand for
\\
\begin{equation}
a_{1} = \alpha r_{0} + w_{0} + w_{1} + \varepsilon_{0}, \qquad
a_{2} = 1 + r_{0} + \varepsilon_{0} \label{a}
\end{equation}
\\
and
\\
\begin{equation}
b_{1} = w_{1} + (\alpha + \textstyle{1\over{3}})\varepsilon_{0},
\qquad b_{2} = \varepsilon_{0}, \label{b}
\end{equation}
\\
respectively.
It follows that
\\
\begin{equation}
\rho_{m}(x) =
\rho_{\varphi}^{(0)}\,(1+r_{0})^{3\beta_{1}}\,x^{-3\beta_2}\,
[r_{0}+\varepsilon_{0}(1-x)]\,[1+r_{0}+\varepsilon_{0}(1-x)]^{-3\beta_{1}}\,,
\label{rhom3}
\end{equation}
as well as
\begin{equation}
\rho_{\varphi}(x) =
\rho_{\varphi}^{(0)}\,(1+r_{0})^{3\beta_{1}}\,x^{-3\beta_{2}}\,[1+r_{0}+\varepsilon_{0}(1-x)]^{-3\beta_{1}},
\label{rhos3}
\end{equation}
\\
with
\[
\beta_{1}=\frac{b_{1}}{b_{2}}-\frac{a_{1}}{a_{2}}=\frac{w_{1}}{\varepsilon_{0}}+
\left(\alpha+\textstyle{1\over{3}}\right)- \left[\frac{\alpha
r_{0}+w_{1}+w_{0}+\varepsilon_{0}}{1+r_{0}+\varepsilon_{0}}
\right],
\]
and
\[
\beta_2=1+\frac{a_1}{a_2}=\frac{1+w_{1}+2\varepsilon_0+r_{0}(1+\alpha)}{1+r_{0}+\varepsilon_{0}}\,.
\]

The Hubble function
\\
\begin{equation}
H(x) = H_{0}
\,(1+r_{0})^{\frac{3\beta_{1}-1}{2}}\,x^{\frac{-3\beta_{2}}{2}}\,
\,[1+r_{0}+\varepsilon_{0}(1-x)]^{\frac{(-3\beta_{1}+1)}{2}} \, ,
\label{hx3}
\end{equation}
\\
follows from the Friedmann's equation.

\begin{figure}[th]
\includegraphics[height=4.5in,width=5.5in,angle=0,clip=true]{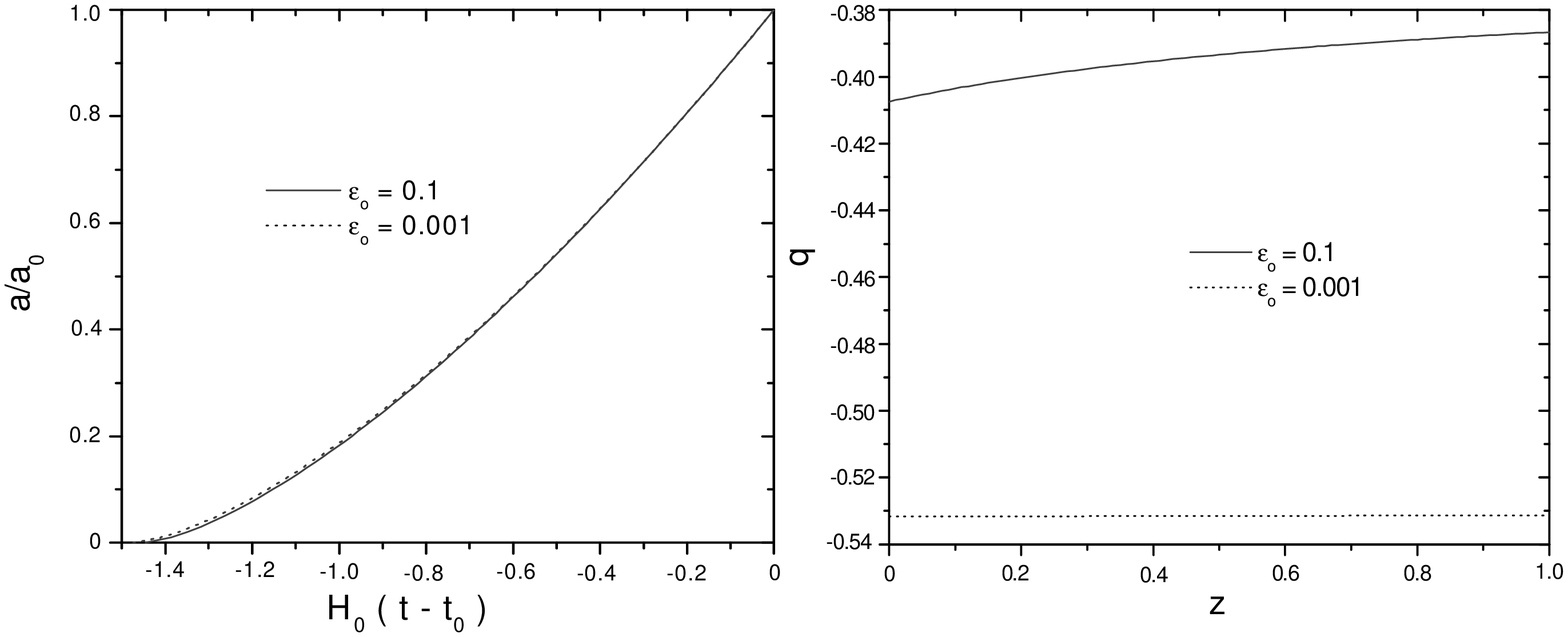}
\vspace{-5.0cm} \caption{The left panel shows the scalar factor as
a function of cosmological time. The right panel shows the
deceleration parameter $q$ as a function of the redshift $z=
x^{-1}-1$. In both cases $w_{0}$=-0.9, $w_{1}$=0.002, $r_{0}=3/7$
and $\alpha$=-0.2.} \label{sq}
\end{figure}

 Although last expression is
comparatively simple, the scale factor derived from it is not
\\
$$
\frac{3\beta_2}{2}(1+r_{0})^{\frac{3\beta_{1}-1}{2}}\,
H_0\,(t-t_{0})=x^{\frac{3\beta_{2}}{2}}(1+r_{0}+\varepsilon_0(1-x))^{\frac{3\beta_{1}-1}{2}}
$$
\begin{equation}
\times\;\,\,_2F_1\left(\left[\frac{3\beta_2}{2},
\frac{1-3\beta_1}{2}\right],\left[ 1+\frac{3\beta_2}{2}
\right];\frac{\varepsilon_0\,x}{1+r_{0}+\varepsilon_0}\right)-C_1\,
, \label{at1}
\end{equation}
where $_2F_{1}$ is the hypergeometric function \cite{hyperg} and
\\
\[
C_1=\left[(1+r_{0})
\left(1-\frac{\varepsilon_0}{1+r_{0}+\varepsilon_{0}}\right)^{-1}\right]^{\frac{3\beta_1-1}{2}}\,_{2}F_{1}(x=1).
\]
\\
Fig. \ref{sq} portrays the evolution of the scale factor in terms
of the cosmological time as well as the deceleration factor
$q\equiv - \ddot{a}/(aH^{2})$ versus the redshift for two
selected values of the parameters.
\\

\begin{figure}[th]
\includegraphics[height=5.5in,width=6.5in,angle=0,clip=true]{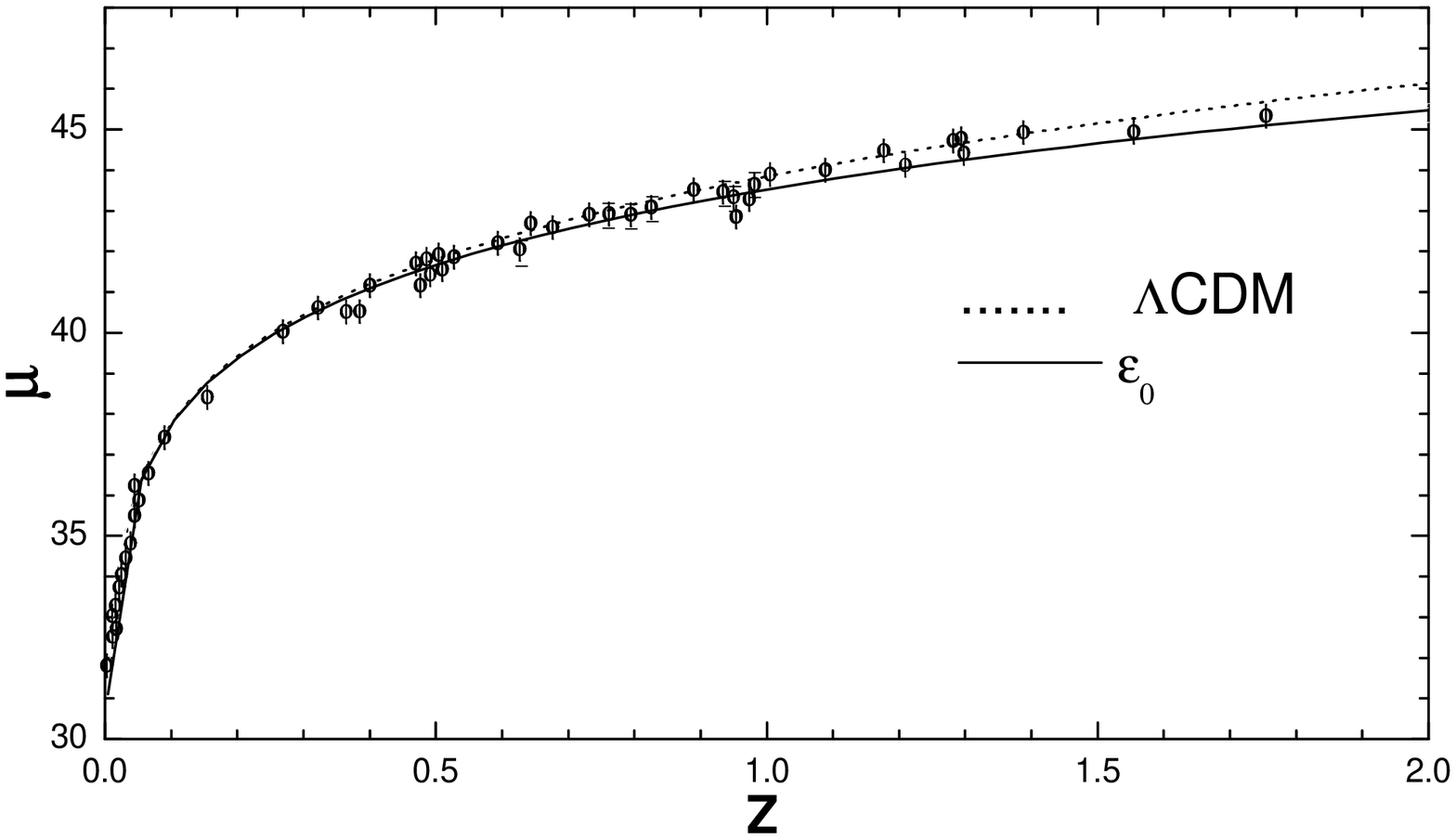}
\vspace{-5.0cm}\caption{Distance moduli vs redshift for the
tachyon--dark matter interacting model. In plotting the graphs the
expression $ \mu = 5 \log d_{L} +25$, with $d_{L} = (1+z)
\int^{z}_{0}H^{-1}(z') dz'$ in units of megaparsecs was assumed.
We have taken the values of the best fit model, namely:
$w_{0}=-0.99$, $w_{1} = 0.95$, $\varepsilon_{0} = 0.0042$, $\alpha
= -0.98$ and $r_{0}=0.136$. For comparison we have also plotted
the prediction of the concordance $\Lambda$CDM model with
$\Omega_{m0} = 0.3$. The data points correspond to the ``gold"
sample of type Ia supernovae of Ref. \cite{adam}.}
\label{vfisupernova}
\end{figure}

\begin{figure}[th]
\includegraphics[height=6.5in,width=6in,angle=0,clip=true]{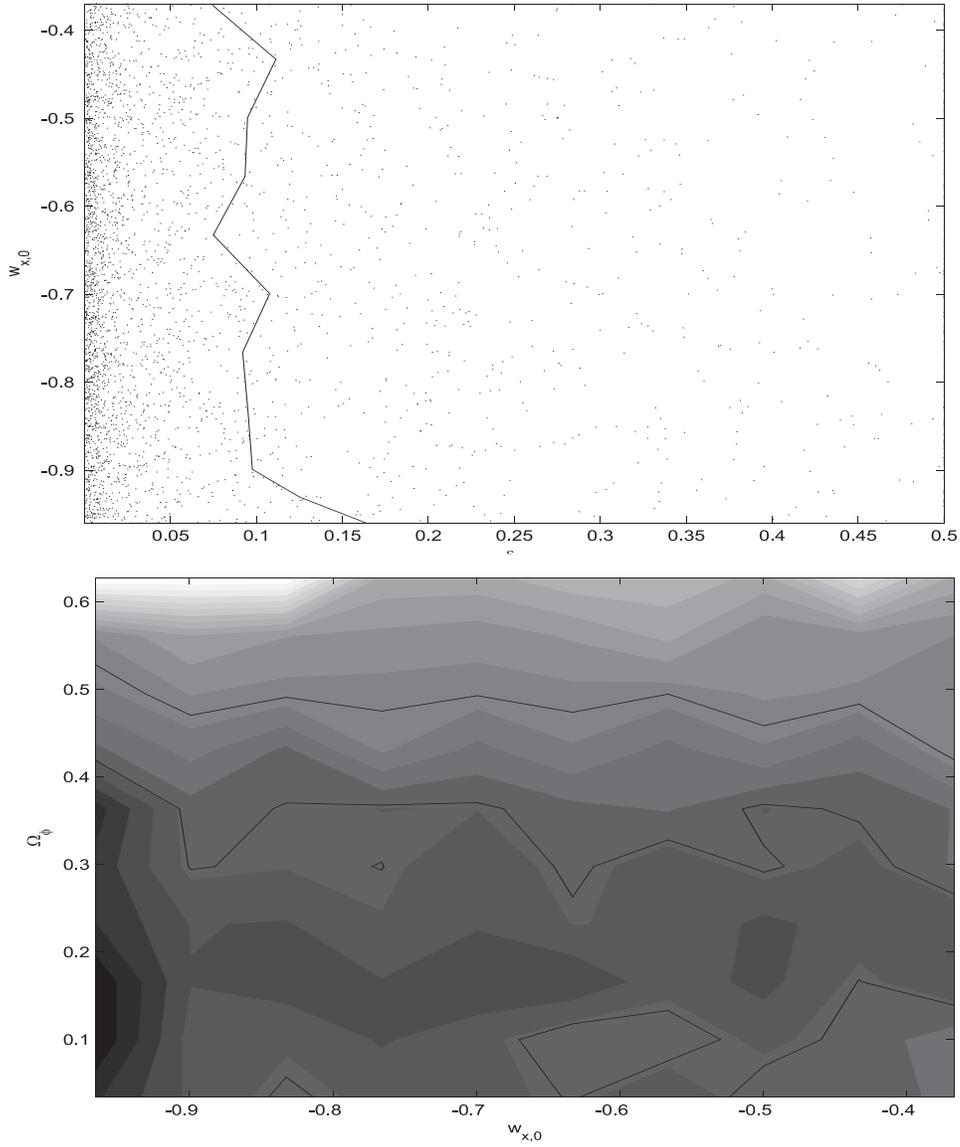}
\vspace{0cm}\caption{Likelihood contours for the tachyon--matter
interacting  model ($w_{x,0}$ vs. $\epsilon_{0}$ -top panel-, and
$\Omega_{\varphi}$ vs. $w_{x,0}$, bottom panel) showing the 68\%
and 98\% confidence intervals. The likelihoods are marginalized
over the rest of parameters.} \label{likesoft2}
\end{figure}

\begin{figure}[th]
\includegraphics[height=6.5in,width=6in,angle=0,clip=true]{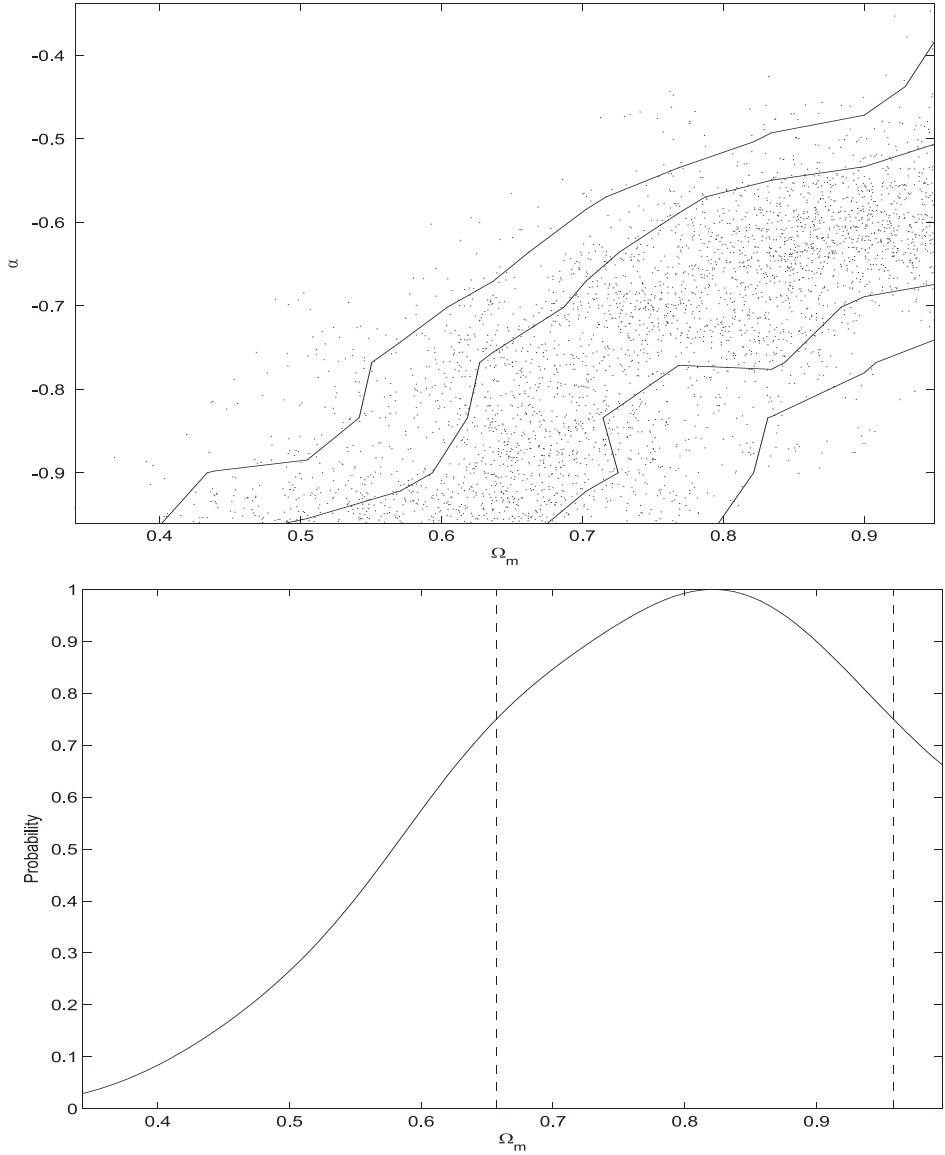}
\vspace{0cm}\caption{Top panel: likelihood contours for the
tachyon--matter interacting  model ($\alpha$ vs. $\Omega_{m})$)
showing the 68\% and 98\% confidence intervals. Bottom panel:
probability function of the matter density parameter. The
likelihoods are marginalized over the rest of parameters.}
\label{likesoft3}
\end{figure}

\begin{figure}[th]
\includegraphics[width=5.0in,angle=0,clip=true]{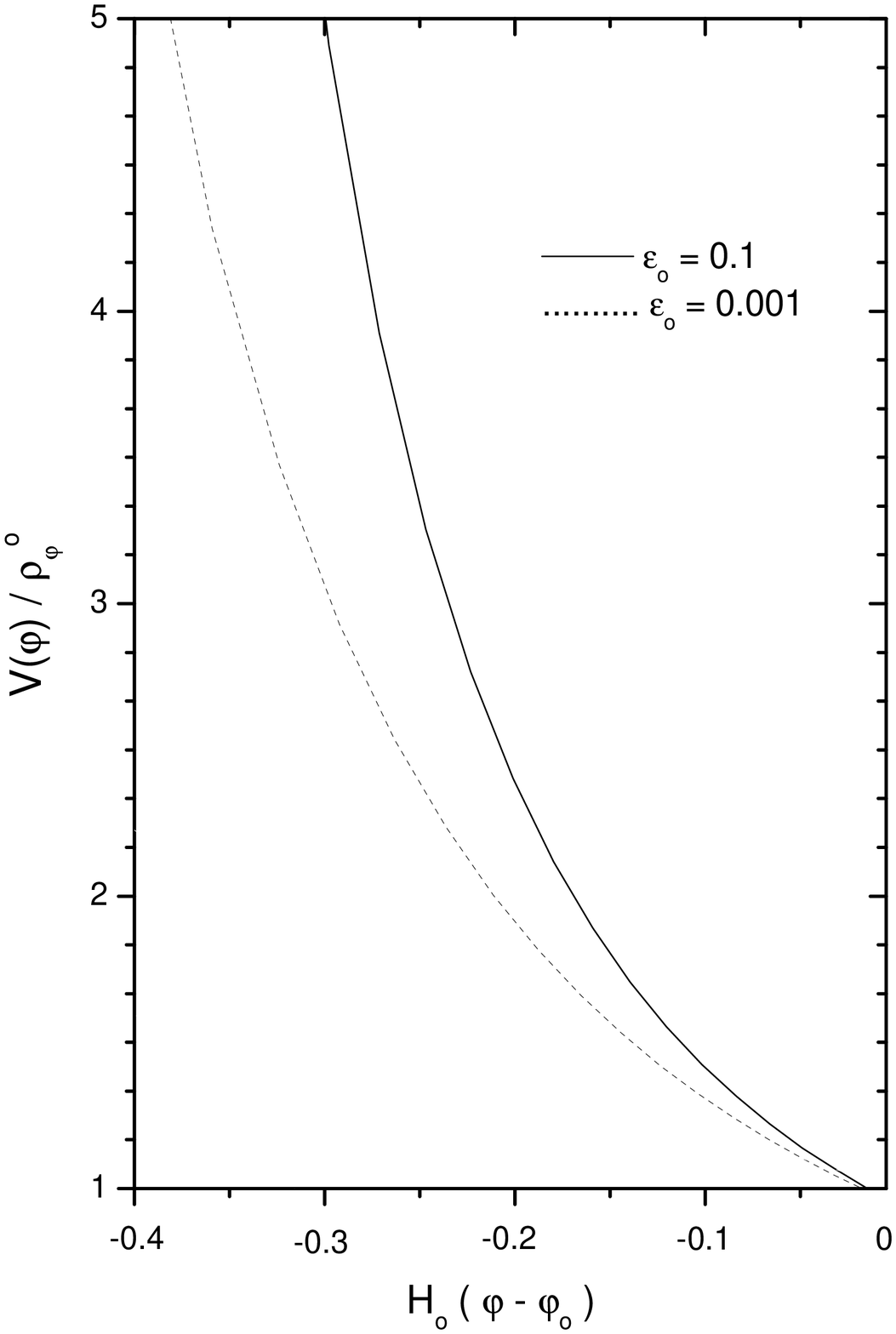}
\caption{The effective potential as a function of the tachyonic
field. Again, $w_{0}$=-0.9, $w_{1}$= 0.002, $r_{0}= 3/7$ and
$\alpha$= - 0.2.} \label{qvfi}
\end{figure}

As Fig. \ref{vfisupernova} shows the model fits the supernova data
points not less well than the concordance $\Lambda$CDM model does.
The likelihood contours, Figs. \ref{likesoft2} and
\ref{likesoft3}, were calculated with the method of Markov's
chains. We used the prior $\Omega_{m} + \Omega_{\vphi} = 1$ and
that the parameters $w_{0}$ and $w_{1}$ are restricted by the
condition that the value of the right hand side of Eq.
(\ref{eta4}) must lay in the interval $[-1,-1/3)$. The mean values
of the parameters are: $\Omega_{\vphi} = 0.246$, $w_{0} = -0.773$,
$w_{1} = 0.22$, $\varepsilon_{0} = 0.0087$, $\alpha = -0.76$.
Here, $\varepsilon_{0}$ is not so weakly constrained by the
supernovae data as in the quintessence model. The present model
predicts a mild evolution of the equation of state parameter with
redshift. This is slightly at variance with the findings of Jassal
{\it et al.} \cite{Jassal}, but agrees with the model independent
analysis of Alam {\it et al.} \cite{alam}.
\\
\\
\\
The interaction term is given by
\noindent
$$
Q(x) =Q_{0}\, \frac{(1+r_{0})^{\frac{9\beta_1-1}{2}}}{[3
r_{0}(\alpha-w_{0})-\varepsilon_{0}]} \left\{3[\alpha-w_{0}-
w_{1}(1-x)][r_{0}+\varepsilon_{0}(1-x)]-\varepsilon_{0}x\right\}
$$
\begin{equation}
\times\,\,x^{\frac{-9\beta_{2}}{2}}\left\{1+r_{0}+\varepsilon_0(1-x)\right\}^{\frac{1-9\beta_{1}}{2}}\,,
\label{delta2}
\end{equation}
with $Q_{0}= \textstyle{1\over{2}}\rho_{\varphi}^{(0)}\,H_{0}[3
r_{0}(\alpha - w_{0})-\varepsilon_{0}]$.
\\
Likewise, the tachyon field and the potential are found to be
\begin{equation}
\varphi(x)=\varphi_{0}+\frac{(1+r_{0})^{\frac{1-3\beta_1}{2}}}{H_0}\int_1^{x}x'^{\frac{3\beta_{2}}{2}-1}
\sqrt{1+w_{0}+w_{1}(1-x')}(1+r_{0}+\varepsilon_0(1-x'))^{\frac{3\beta_1-1}{2}}dx'
\label{campo2}
\end{equation}
and
\begin{equation}
V(x) =
\rho_{\varphi}^{(0)}(1+r_{0})^{3\beta_{1}}\sqrt{-w_{0}- w_{1}(1-x)}\,
x^{-3\beta_{2}} [1+r_{0}+\varepsilon_{0}(1-x)]^{-3\beta_{1}}\, ,
\label{Vx2}
\end{equation}
respectively -see Fig. \ref{qvfi}.

\section{Concluding remarks}
We have studied two models of late acceleration by assuming $(i)$
that dark energy and non-relativistic dark matter do not conserve
separately but the former decays into the latter as the Universe
expands, and $(ii)$ that the present ratio of the dark matter
density to dark energy density varies slowly with time, i.e., $\mid%
\dot{r}\mid_{0} \, \leq H_{0}$. This second assumption is key to
determine the interaction $Q$ between both components.

In the quintessence model (section II) we have considered the
equation of state parameter constant while in the tachyon field
model (section III) we have allowed it to vary slightly. Actually,
there is no compelling reason to impose that this parameter should
be a constant. However, Jassal {\it et al.} \cite {Jassal} have
pointed out that the WMAP data \cite{spergel} imply that in any
case it cannot vary much. By contrast, Alam {\it et al.} using the
sample of ``gold" supernovae of Riess {\it et al.} \cite{adam}
find a clear evolution of $w$ in the redshift interval $0\leq z
\leq 1$; however when strong priors on $\Omega_{m0}$ and $H_{0}$
are imposed this result weakens. Nevertheless, the analysis of
these two papers assume that the two main components (matter and
dark energy) do not interact with each other except
gravitationally. The parameter $w_{0}$ presents degeneration in
both models, therefore we must wait for further and more accurate
SNIa data, perhaps from the future SNAP satellite, or to resort to
complementary observations of the CMB.

In both cases (quintessence and tachyon), we have found analytical
expressions for the relevant quantities (i.e., the scale factor,
the field and the potential) and the solutions are seen to
successfully pass the magnitude-redshift supernovae test -see
Figs. \ref{fisupernova} and \ref{vfisupernova}. Nevertheless, it
is apparent that the the tachyon model favors  rather high values
of the matter density parameter (see bottom right panel of Fig.
\ref{likesoft2}) which is at variance with a variety of
measurements of matter abundance at cosmic scales \cite{peebles}
which, taken as a whole, hint that $\Omega_{m}$ should not exceed
$\sim 0.45$. In consequence, the quintessence model appears
favored over the tachyon model. Our work may serve to build more
sophisticated models aimed to simultaneously account for the
present acceleration and the coincidence problem.

Previous studies of interacting dark energy aimed to solve the
coincidence problem by demanding that the ratio $r$ be strictly
constant at late times needed to prove the stability of $r$  at
such times. This was achieved by showing that the models
satisfied an attractor condition that involved the equation of
state parameter of matter and dark energy as well as the Hubble
factor and its temporal derivative \cite{plb,interacting,dw}.
Since in the case at hand the coincidence problem is solved with a
(slowly) varying ratio $r$ no stability proof is necessary at all
and no attractor condition is needed.

Our analysis was confined to times not far from the present (i.e.,
for $x \sim {\cal O}(1)$. To recover the evolution of the Universe
at earlier times ($x \ll 1$), when the matter density dominated
and produced via gravitational instability the cosmic structures
we observe today, we must generalize our study along the lines of
Refs. \cite{interacting} and \cite{dw} and include the baryon
component in the dynamic equations as an uncoupled fluid.

We restricted ourselves to scenarios satisfying $w > -1$.
Scenarios with $w <- 1$ (the so-called ``phantom" energy models)
violate the dominant energy condition though, nevertheless, they
are observationally favored rather than excluded \cite{alessandro}
and exhibit interesting features \cite{caldwell} that might call
for ``new physics". We defer the study of  phantom models
presenting soft coincidence to a future publication.

\acknowledgments Thanks are due to Germ\'{a}n Olivares for his
computational assistance. DP is grateful to the ``Instituto de
F\'{\i}sica de la PUCV" for financial support and kind
hospitality. SdC was supported from Comisi\'on Nacional de
Ciencias y Tecnolog\'{\i}a (Chile) through FONDECYT grants N$^0$s
1030469, 1010485 and 1040624 as well as by PUCV under grant
123.764/2004. This work was partially supported by the old Spanish
Ministry of Science and Technology under grant BFM--2003--06033,
and the ``Direcci\'{o} General de Recerca de Catalunya" under
grant  2001 SGR--00186.

\end{document}